# On Service-Chaining Strategies using Virtual Network Functions in Operator Networks

Abhishek Gupta, M. Farhan Habib, Uttam Mandal, Pulak Chowdhury, Massimo Tornatore, and Biswanath Mukherjee

**Abstract**—Network functions (e.g., firewalls, load balancers, etc.) have been traditionally provided through proprietary hardware appliances. Often, hardware appliances need to be hardwired back-to-back to form a service chain providing chained network functions. Hardware appliances cannot be provisioned on-demand since they are statically embedded in the network topology, making creation, insertion, modification, upgrade, and removal of service chains complex, and also slowing down service innovation. Hence, network operators are starting to deploy Virtual Network Functions (VNFs), which are virtualized over commodity hardware. VNFs can be deployed in Data Centers (DCs) or in Network Function Virtualization (NFV)-capable network elements (nodes) such as routers and switches. NFV-capable nodes and DCs together form a Network-enabled Cloud (NeC) that helps to facilitate the dynamic service chaining required to support today's evolving network traffic and its service demands. In this study, we focus on the VNF service chain placement and traffic routing problem, and build a model for placing a VNF service chain while minimizing network-resource consumption. Our results indicate that a NeC having a DC and NFV-capable nodes can significantly reduce network-resource consumption.

✦

## 1 INTRODUCTION

TODAY'S communication networks comprise of a large variety of proprietary hardware appliances (e.g., middle-boxes (MBs)) which support network functions such as firewalls, Network Address Translators (NATs), Quality-of-Service (QoS) analyzers, etc. Often, these hardware appliances need to be traversed in sequence, forming a service chain, which provides chained network functions to specific traffic flows. As hardware-based functions are embedded in the network topology, this static allocation enforces topological constraints on network traffic, which is required to pass through a set of specific nodes to satisfy service requirements. Here, service requirements refer to the service required by the traffic, which depends on traffic type. For example, video traffic in a network requires video-optimization service. The operator provisions the video-optimization service[1] using a video optimizer and a firewall. For the video traffic to avail this service, it is required to traverse the firewall and then the video optimizer. Sequential traversal of the units (here, firewall and video optimizer) has the effect of video-optimization service being deployed for the video traffic.

The problem of satisfying service requirements is escalating as, with new bandwidth-intensive cloud applications becoming popular,

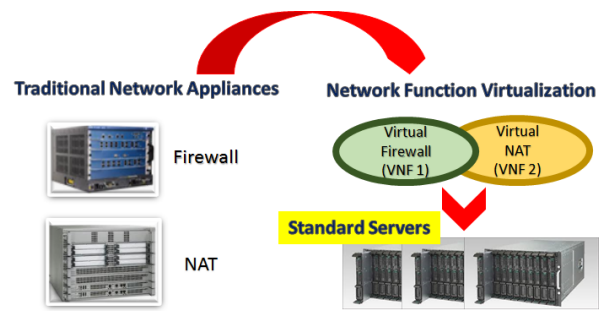

Fig. 1: Network Function Virtualization (NFV) approach.

operators and enterprises functioning as operators[2] are required to provide more network services to their clients. Current networks support these services with vendor-specific hardware appliances, which are difficult to configure, modify, and upgrade, so the cycle of service introduction, modification, and upgrade/removal is becoming more complex. This problem is compounded because MBs require frequent upgrades due to rapid innovation in technology and increase in traffic volume. This complexity leads operators to route all traffic through the chain, irrespective of service requirements, to avoid misconfiguration and reduce downtime [1].

Network Function Virtualization (NFV) [2] provides the operator with the right tools to handle network traffic more effectively and dynamically. As shown in Fig. 1, the predominant idea behind NFV is to replace vendor-specific hardware with Commercial-Off-The-Shelf (COTS) hardware such as servers, switches, and storage [3], which are placed in Data Centers (DCs) or, more generally, in network nodes equipped with server capabilities. This leads to more

- A. Gupta, P. Chowdhury, M. Tornatore and B. Mukherjee are with the University of California, Davis, USA.
  E-mail: {abgupta, pchowdhury, mtornatore, bmukherjee}@ucdavis.edu
- M. Tornatore is also with Politecnico di Milano, Italy.
  E-mail: tornator@elet.polimi.it
- M. F. Habib is with Intel Corporation, Folsom, USA.
  E-mail: mohammad.f.habib@intel.com
- U. Mandal is with Cisco Inc., San Jose, USA.
  E-mail: utmandal@cisco.com

1. Video-optimization service can be provisioned using different functions. Here, we assume it is provided by chaining a firewall and a video optimizer.

2. For example, a university which needs to provide a live-streaming service.

flexibility in deployment of services; hence, service innovation becomes easier. Operators can scale the service according to traffic intensity while also generating more innovations and revenue. It is for this flexibility in deployment that VNF-based service chains are being deployed by operators. However, NFV realization has major challenges in key areas of performance, management, security, scalability, resiliency, and reliability [4] (with line-rate performance of VNFs and their orchestration being among the most significant). These challenges stem from the necessity of having a "carrier-grade" NFV infrastructure, as network operators are bound by more stringent service-level agreements (SLA) and regulations than a typical enterprise.

In this work, we focus on providing a quantitative measure for operators to understand the reductions in Operating Expenditure (OPEX) (i.e., network resource consumption) to be expected, given the Capital Expenditure (CAPEX) (i.e., number of network nodes capable of hosting VNFs) on NFV platforms. Here, we assume the network resource consumption to be the main contributor to OPEX because the current DC-based concept of NFV will lead to frequent redirection of traffic, leading to more bandwidth being consumed. Operators with their limited bandwidth installation will be hard-pressed to provide network resources for this frequent redirection. Thus, efficient bandwidth utilization is a major priority for operators. We study strategies for service chaining, and develop a mathematical model for optimal placement of VNFs which minimizes the network-resource consumption while ensuring that traffic in the network meets its service requirements. We compare the network-resource consumption across the service-chaining strategies based on our model, and provide a quantitative estimate on the reduction of network-resource consumption achieved when the NFV infrastructure contains network nodes capable of supporting NFV besides the centralized NFV infrastructure, e.g., DCs.

The rest of the study is organized as follows. First, we review the relevant literature on VNF placement in Section 2. In Section 3, we discuss CPU-core-to-throughput relationship of VNFs and elaborate on the concept of a Network-enabled Cloud (NeC). In Section 4, the concept of service chaining is introduced, and various service-chaining strategies are discussed. In Section 5, we describe the VNF service chain placement problem and the model to optimize the network-resource consumption. Section 6 presents results for various service-placement strategies. Concluding remarks are provided in Section 7.

## 2 RELATED WORK

The problem to evaluate the impact of various VNF-based service-chaining strategies on network-resource consumption follows from a combination of traffic routing (multicommodity flow problem) and VNF placement (location problem). A number of studies have appeared recently on this problem. In [5], the objective is to reduce the number of servers deployed for VNFs using an Integer Linear Program (ILP) but it does not account for service chaining of VNFs explicitly. Ref. [6] studies specification and placement of VNF service chains. It develops a heuristic to specify the VNF service chain and a Mixed Integer Quadratically Constrained Program (MIQCP) for the VNF placement problem. In [7], the VNF placement and routing problem is modeled as a Mixed Integer Linear Program (MILP) to place services optimally for flows and minimize network-resource consumption, and heuristics are developed to place services optimally for a large number of flows.

Ref. [8] also solves the problem of VNF service chain placement using an MILP and gives insights into trade-offs between legacy and NFV-based traffic engineering. In [9], the authors determine the number of VNFs required and their placement to optimize OPEX while adhering to SLAs using an ILP, while heuristics based on dynamic programming are used to solve larger instances of the problem. Ref. [10] also models the problem using an ILP to reduce the end-to-end delays and minimize resource over-provisioning while providing a heuristic to do the same. Ref. [11] considers the enterprise WAN scenario same as our work; however, it does not account for the CPU core requirements of VNFs and proposes a greedy heuristic while our model computes optimal values. In [12], the authors reduce the cost of operation using an ILP and devise a greedy algorithm for the same. Ref. [13] gives a scalable orchestrator which can scale services (service chain of VNFs) depending on demand. Ref. [14] looks at dynamic scaling of services depending on the workload while minimizing operational cost of cloud service providers using both static and dynamic approaches. Ref. [15] models the optimal deployment of a service chain as a MILP while optimizing host and bandwidth resources.

The dynamic (on-line) placement of VNFs is required to account for rapidly-changing service demands of traffic. Ref. [16] proposes an on-line optimization algorithm for joint path selection and placement of VNFs using SDN and NFV. Refs. [17] and [18] propose algorithms for online VNF scheduling; while [18] accounts for service chaining, [17] does not. Ref. [19] proposes a deterministic online algorithm for VNF placement with proven upper and lower bounds for computation time. Ref. [20] looks at on-line placement of VNFs by designing a VNF orchestration architecture, while Refs. [21] and [22] demonstrate on-line VNF placement using Openflow and Openstack, respectively.

In this work, we develop a mathematical model for VNF service chain placement and traffic-flow routing with the objective of minimizing network-resource consumption. This model is used to study the network-resource consumption by comparing different service-chaining strategies. While employing a similar approach (ILP) as related works, our model is the first, to the best of our knowledge, to account for different deployment strategies for service chaining, allowing us to compare different strategies (e.g., by changing the number of nodes that can host VNFs).

## 3 VIRTUAL NETWORK FUNCTION (VNF) AND NETWORK-ENABLED CLOUD (NEC)

Virtual network functions (VNFs) are software modules that abstract hardware-based functions and are run as virtual machines (VMs). The requirements of a VNF are similar to those of a VM in terms of computing and storage resources, i.e., both require CPU cores and memory (RAM/hard disk). However, traditional VMs are enterprise applications (e.g., database applications) deployed in cloud-computing environments while VNFs abstract network functions which process network traffic at line rate. This makes VNFs more bandwidth-intensive (virtualized routers) and compute-intensive (virtualized firewall: computational overhead for per-packet processing) while enterprise application VMs are more memory-intensive and compute-intensive.

Although a VNF instantiation requires a certain amount of memory and disk space, the performance of VNF scaling will depend on the CPU-core-to-throughput relationship. This relationship will depend on the VNF type which can be seen from

| Application | Throughput | | |
|---|---|---|---|
| | 1 Gbps | 5 Gbps | 10 Gbps |
| NAT | 1 CPU | 1 CPU | 2 CPUs |
| IPsec VPN | 1 CPU | 2 CPUs | 4 CPUs |
| Traffic Shaper | 1 CPU | 8 CPUs | 16 CPUs |

TABLE 1: VNF requirements as per throughput [23].

Table 1, where, e.g., a NAT (Network Address Translator) performs basic IP addressing functions, making it less CPU-intensive, while a Traffic Shaper needs to identify application traffic and perform operations which are compute-intensive, and results in a large number of CPU cores being used for higher throughput. So, the CPU-core-to-throughput relationship is an essential characteristic of VNF operation, and it is the basis for VNF allocation in our mathematical model, which is explained in Section 5.

Since VNFs are essentially VMs, and they can be deployed on any hardware having VM support, DCs and/or network elements equipped with computing and storage resources can be part of the NFV infrastructure and are called "NFV-capable". These VNFs can be deployed at edge routers, DCs, central offices, mobile switching centers, local points of presence, or customer premises using commodity hardware/servers. An edge router, e.g., Juniper MX Series, includes the JunOS VApp engine [24] for hosting VMs on the router itself. Also, virtual routers like the Cisco 1000V series can be run on top of servers. This enables the creation of a Network-enabled Cloud (NeC) to host VNFs, as shown in Fig. 2.

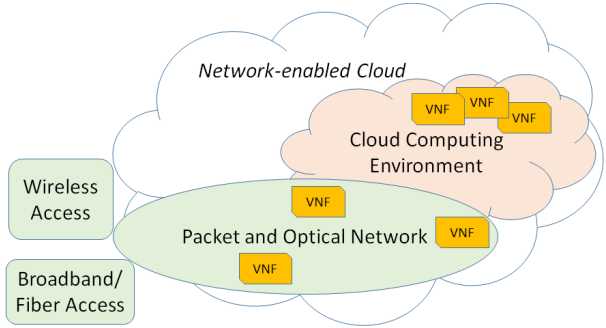

Fig. 2: Network-enabled cloud (NeC).

Cloud computing provides elastic computing and storage facilities where resources can be provisioned on demand and is traditionally restricted to DCs. Today, DCs are the first solution for service chaining. But, if all the VNFs are hosted in a DC and all traffic is forwarded through the DC, the network-resource consumption may increase heavily. As emerging services often need resources residing in different cloud or network domains, a NeC [25] would facilitate provisioning of resources not only in DC but also in other elements of the associated network infrastructure (network nodes), making VNF deployment across the network more dynamic and utilization of network resources more efficient.

# 4 SERVICE CHAINING

Network functions process traffic either singularly or in sync with other network functions, forming a "service chain". The term "service chaining" is used "to describe the deployment of such functions, and the network operator's process of specifying an ordered list of service functions that should be applied to a deterministic set of traffic flows" [26]. Therefore, a "service chain" means a set of network functions placed in a specific order. Traffic flows are classified, and depending on the service required, they pass through a specific service chain. This classification can be done on per-port basis, per-subscriber basis, or on the basis of location (in the network), among others. Therefore, operators will have different routing policies according to the traffic type, and route the traffic through the service chain to satisfy service requirements.

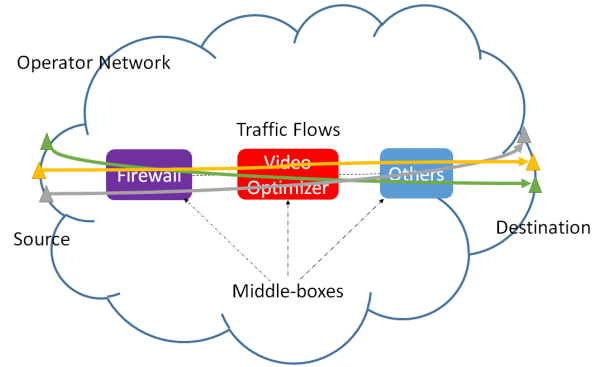

Fig. 3: Static service chaining.

An example of a "service chain" is shown in Fig. 3. This service chain is configured as static middle-boxes (MBs), wired together. With rapid increases in traffic volume, traffic variety, and service requirements, operators are looking at VNFs for a more flexible method of service chaining.

## 4.1 Service-Chaining (SC) Strategies

We now describe various service-chaining strategies characterized by varying degree of flexibility in deployment of service chains.

*SC using Middle-Boxes (MB):* This is the traditional way to create a service chain. Each network function is a hardware-based appliance (MB), and a service chain is formed by hardwiring a sequence of MBs.

*SC using Data Center (DC):* Here, the network functions are deployed in the form of VNFs at DCs. In this scheme, all service chains will be deployed inside the DC, therefore all traffic will have to be routed to the DC for service.

*SC using DC plus 'x' NFV-capable nodes (DC NFV x):* In this scheme, service chains can be implemented both inside the DC and/or in a set of NFV-capable nodes. A particular case is the "DC NFV ALL" strategy, where all nodes are NFV-capable. Although having all network nodes as NFV-capable can be cost-prohibitive (this would be an "ideal NeC" case), it forms a good benchmark for comparison of results. In particular, if core count is not a limiting factor, traffic flows will traverse their service chain on the shortest path, and will result in minimal resource consumption across all the scenarios. The "DC NFV ALL" is our best-case strategy.

*SC using NFV:* This is a completely-distributed NFV scenario where all nodes are NFV-capable and there is no centralized NFV infrastructure like a DC. We refer to this scenario as "NFV ALL" since all nodes are NFV-capable. Here too, if core count is not a limiting factor, traffic flow will follow the shortest path and resource consumption will be minimal (same as "DC NFV ALL").

The major difference between "DC NFV ALL" and "NFV ALL" is that "DC NFV ALL" has a centralized NFV infrastructure like a DC while "NFV ALL" is a completely-distributed scenario where no DC exists.

# 5 PROBLEM DESCRIPTION

A network can support multiple types of traffic, and each type of traffic requires service. The service is accomplished by making the traffic traverse a service chain of VNFs deployed for that service. Each service requirement can be satisfied by varying the configurations of the service chain, i.e., the same service can be delivered by building the service chain using different VNFs. For example, if a service chain requires a firewall and a NAT, the service chain can be deployed as comprising of two VNFs (firewall and NAT); or a single VNF consisting of both the firewall and the NAT functionality can be deployed. The type of VNF deployed depends on the VNF vendor and the network operator. The configuration of service chains depending on the service required is an important problem but is out of scope of our current work[3]. The operator needs to satisfy the service requirements for all the traffic flows using minimal network resources. Therefore, we model the problem as an optimization problem where the objective is to minimize the network-resource consumption by optimally placing VNF service chains. This model formulates and solves the problem that has to be realized in practice by the NFV management platforms.

## 5.1 Problem Statement

Given a network topology, capacity of the links, a set of DC locations, a set of network nodes with virtualization support (NFV-capable nodes), traffic flows between source-destination pairs, bandwidth requirement of the traffic flows between the source-destination pairs, set of network functions required, and the service chain, we determine the placement of VNFs to minimize network-resource (bandwidth) consumption.

## 5.2 Input Parameters

- $G(V, E)$: Physical topology of the network; $V$ is set of nodes and $E$ is set of links.
- $V_{DC} \subset V$: Set of DC locations.
- $V_{NF} \subset V$: Set of NFV-capable nodes.
- $\Psi_{(s,d)}$: Set of source $s \in V$ and destination $d \in V$ pairs requesting traffic flows among them.
- $\Phi_{(s,d)}$: Traffic request from source $s$ to destination $d$.
- $K_{s,d}$: $K$ shortest paths from source $s$ to destination $d$.
- $\Gamma$: Set of network functions.
- $\Pi$: Service chain of functions.
- $R_{(s,d)}^{(i,j)}$: Set of paths from source $s$ to destination $d$ passing through link $(i, j) \in E$.
- $\theta$: Number of cores in a NFV node.
- $N_f$: Number of cores required by function $f$ to serve a unit of throughput.
- $\Upsilon$: Memory in a NFV node.
- $\varsigma_f$: Memory required by function $f$ irrespective of throughput.
- $\chi_f$: Memory required by function $f$ to serve a unit of throughput.
- $L_{s,d}^p$: Length of path $p$ between source $s$ and destination $d$.
- $C_{i,j}$: Bandwidth capacity of link $(i, j) \in E$.

3. Here, we assume that the operator has the technical capability to configure VNF service chains for the required service.

- $S_{u,v}^p$: Set of node pairs $(u, v)$ such that $u$ and $v$ are nodes on path $p$ and $u \in V_{NF}$[4] occurring before $v \in V_{DC} \cup V_{NF}$.

## 5.3 Variables

- $r_p^{(s,d)} \in \{0, 1\}$: 1 if path $p$ is chosen between source $s$ and destination $d$.
- $l_v^f \in \{0, 1\}$: 1 if function $f$ is used at node $v$.
- $q_{p,(s,d)}^{f,v} \in \{0, 1\}$: 1 if function $f$ is located in node $v$ of path $p$ between $(s, d)$.
- $j_{p,(s,d),(u,v)}^{f_1, f_2} \in \{0, 1\}$: 1 if functions $f_1$ and $f_2$ occur in service chain order at nodes $u$ and $v$ of path $p$ between $(s, d)$.

## 5.4 Problem Formulation

We mathematically formulate the problem as an Integer Linear Program (ILP), as follows.

$$Minimize: \qquad \sum_{(s,d) \in \Psi_{s,d}} \sum_{p \in K_{s,d}} r_p^{(s,d)} \times L_{s,d}^p \times \Phi_{s,d} \qquad (1)$$

such that

$$\sum_{p \in K_{s,d}} r_p^{(s,d)} = 1 \qquad \forall (s,d) \in \Psi_{(s,d)} \qquad (2)$$

$$\sum_{(s,d) \in \Psi_{s,d}} \sum_{r_p^{(s,d)} \in R_{(s,d)}^{(i,j)}} r_p^{(s,d)} \times \Phi_{s,d} \le C_{(i,j)} \qquad \forall (i,j) \in E \qquad (3)$$

$$\sum_{(s,d) \in \Psi_{s,d}} \sum_f l_{v,(s,d)}^f \times \Phi_{s,d} \times N_f \le \theta \qquad \forall v \in V_{NF} \qquad (4)$$

$$\sum_{(s,d) \in \Psi_{s,d}} \sum_f l_{v,(s,d)}^f \times \varsigma_f \le \Upsilon \qquad \forall v \in V_{NF} \qquad (5)$$

$$\sum_{(s,d) \in \Psi_{s,d}} \sum_f l_{v,(s,d)}^f \times \Phi_{s,d} \times \chi_f \le \Upsilon \qquad \forall v \in V_{NF} \qquad (6)$$

$$q_{p,(s,d)}^{f,v} = l_{v,(s,d)}^f \cap r_p^{(s,d)} \qquad \forall (s,d) \in \Psi_{(s,d)}, \\ \forall f \in \Gamma, \quad \forall p \in K_{s,d}, \\ \forall v \in p \mid v \in V_{DC} \cup V_{NF} \qquad (7)$$

$$\sum_{p \in K_{s,d}} \sum_v q_{p,(s,d)}^{f,v} \ge 1 \qquad \forall f \in \Gamma, \\ \forall (s,d) \in \Psi_{(s,d)}, \\ \forall v \in p \mid v \in V_{DC} \cup V_{NF} \qquad (8)$$

$$j_{p,(s,d),(u,u)}^{(f_1,f_2)} \ge q_{p,(s,d)}^{f_1,u} \qquad \forall (s,d) \in \Psi_{(s,d)}, \\ \forall p \in K_{s,d}, \quad \forall u \in p \mid u \in V_{DC}, \\ \forall (f_1, f_2) \in \Gamma \mid (f_1 \to f_2) \in \Pi \qquad (9)$$

$$j_{p,(s,d),(u,v)}^{(f_1,f_2)} = q_{p,(s,d)}^{f_1,u} \cap q_{p,(s,d)}^{f_2,v} \qquad \forall p \in K_{s,d}, \\ \forall (f_1, f_2) \in \Gamma \mid (f_1 \to f_2) \in \Pi, \\ \forall (s,d) \in \Psi_{(s,d)}, \quad \forall (u,v) \in S_{u,v}^p \qquad (10)$$

4. We assume that, if $u \in V_{DC}$, then the subsequent VNFs of the service chain will also be deployed in the DC, i.e., we will only have one node $u$, not a pair of nodes $(u, v)$. Hence, $u \in V_{NF}$.

$$\sum_{p \in K_{s,d}} \sum_{(u,v) \in p} j_{p,(s,d),(u,v)}^{(f_1,f_2)} \geq 1 \qquad \forall (u,v) \in S_{u,v}^p,$$

$$\forall (s,d) \in \Psi_{(s,d)}, \tag{11}$$

$$\forall (f_1, f_2) \in \Gamma \mid (f_1 \to f_2) \in \Pi$$

$$\sum_t j_{p,(s,d),(t,u)}^{(f_1,f_2)} \geq j_{p,(s,d),(u,v)}^{(f_2,f_3)} \qquad \forall p \in K_{s,d},$$

$$\forall (s,d) \in \Psi_{(s,d)}, \tag{12}$$

$$\forall (f_1, f_2, f_3) \in \Gamma \mid (f_1 \to f_2 \to f_3) \in \Pi,$$

$$\forall (t,u,v) \in p \mid ((t,u) \in V_{NF}, v \in V_{DC} \cup V_{NF})$$

$$\sum_{(u,v)} j_{p,(s,d),(u,v)}^{(f_1,f_2)} \leq 1 \qquad \forall p \in K_{s,d},$$

$$\forall (s,d) \in \Psi_{(s,d)}, \tag{13}$$

$$\forall (f_1, f_2) \in \Gamma \mid (f_1 \to f_2) \in \Pi$$

The objective function in Eq. (1) calculates the total bandwidth consumed by all the requested source-destination traffic flows. We enforce that traffic between a source-destination pair is served by a single path using Eq. (2), i.e., source-destination traffic is not split among multiple paths. Amount of flows that can be provisioned on a link is constrained by the bandwidth of the link (Eq. (3)).

Each flow has service requirements which need to be satisfied. We deploy VNFs for this purpose. Each VNF, depending on the application it virtualizes, requires a certain number of CPU cores for processing a unit of throughput (in terms of bandwidth). The CPU core requirement is not a constraint in a DC setting. However, in case of a NFV-capable node, computation power is limited, which will impact the assignment of VNFs at that node. This constraint is realized using Eq. (4). Based on the chain deployed for the service, the flow might be processed by one or more VNFs.

For certain VNFs, memory requirement will be more critical than CPU cores. Eq. (5) is for VNFs which have static memory requirements while Eq. (6) describes VNFs whose memory requirements scale with traffic. We consider that all the VNFs in the service chain will be of one type or the other. These equations are only enforced for the results discussed in Section (6.3).

To provide required service to traffic, it has to be mandated that it traverses the VNFs in strict order as described by the service chain. Eqs. (7) to (13) ensure that the service chain is implemented and traversed in the correct sequence for the traffic flows.

Eq. (7)[5] checks if a VNF is located on the path taken by the traffic flows, while Eq. (8) enforces that this VNF exists on one of the paths, so that there is one path to select. If a particular VNF is available on a path, we have to enforce that the next VNF in the service chain is also available on that path. Eq. (9) enforces service chaining in a DC, i.e., if a function is located at a DC, then all its successors in the service chain will also be located here. This is logical as a DC has sufficient resources to deploy many VNFs. In the case of a NFV-capable node, we need to ensure that the successive function is in the path either at the same node or at a different successive NFV-capable node (part of the path), and this is enforced using Eq. (10)[6], while Eq. (11) enforces that the constraint specified by Eq. (10) is realized in at least one of the paths between the source and the destination. Eq. (12) enforces service chaining inside a network node by constraining that a later dependency ($f_2 \to f_3$) is possible only if an earlier ($f_1 \to f_2$) dependency is satisfied. Eq. (13) enforces that the dependencies are enforced exactly in the path chosen.

Note that, by changing the parameters of the ILP formulation, we can model all the SC strategies described in Section 4.

The "MB" strategy is implemented by replacing Eq. (4) with a location constraint which constrains the number of MBs that can be placed at a node. "DC" strategy is modeled by enforcing that $V_{NF} = \phi$ while "DC NFV x" strategy is modeled by increasing the number of nodes in $V_{NF}$ to $x$. Finally, the "NFV ALL" strategy involves setting $V_{DC} = \phi$.

Note that a network may have multiple traffic flows generated from different application-layer programs. Each traffic type will have its own service requirements, each of which may be satisfied by different service chains implementing the same service. In this study, we try to reduce simulation time by considering the service chain to be given and one type of traffic in the network in any instance.

# 6 ILLUSTRATIVE NUMERICAL EXAMPLES

We first run instances of our optimization model on an Enterprise WAN running over the NSFNet topology (see Fig. 4(a) showing the headquarter (HQ) and branch offices). Each network link has 40 Gbps capacity in each direction. Bidirectional traffic flows are generated between HQ and branch offices as shown in Fig. 4(b), and flows going to and from the HQ carry 1.5 times the traffic between branch offices. Average value of each traffic flow can assume increasing values from 1 Gbps to 10 Gbps. Traffic flows will take different paths depending on the locations of the MBs, DCs, and NFV-capable nodes. We consider only one type of traffic (i.e., one type of service chain) in the network at any instance.

Figure 4(c) shows two possible service chains that are pertinent to an Enterprise WAN scenario, and can be deployed between a HQ and a branch office inspired from the use cases discussed in [27]. Service chain (SC) 1 provides Network Address Translation (NAT) for outgoing traffic, after which Traffic Shaping (TS) is used to direct application traffic for Application Optimization (AO), followed by encryption (IPSec), and finally a WAN Acelerator (WANA). Similarly, SC2 provides a firewall for filtering incoming traffic, followed by an Intrusion Detection System (IDS) for

$$\forall f \in \Gamma, \quad \forall (s,d) \in \Psi_{(s,d)}, \quad \forall p \in K_{s,d}, \quad \forall v \in p \mid v \in V_{DC} \cup V_{NF}$$

Eq. (7) can be linearly represented as below.

$$q_{p,(s,d)}^{f,v} \leq l_{v,(s,d)}^f \tag{14}$$

$$q_{p,(s,d)}^{f,v} \leq r_p^{(s,d)} \tag{15}$$

$$q_{p,(s,d)}^{f,v} \geq l_{v,(s,d)}^f + r_p^{(s,d)} - 1 \tag{16}$$

$$\forall (s,d) \in \Psi_{s,d}, \quad \forall p \in K_{s,d}, \quad \forall (u,v) \in S_{u,v}, \quad \forall (f_1,f_2) \in \Gamma \mid (f_1 -> f_2) \in \Pi$$

Eq. (10) can be linearly represented as below.

$$j_{p,(s,d),(u,v)}^{(f_1,f_2)} \leq q_{p,(s,d)}^{f_1,u} \tag{17}$$

$$j_{p,(s,d),(u,v)}^{(f_1,f_2)} \leq q_{p,(s,d)}^{f_2,v} \tag{18}$$

$$j_{p,(s,d),(u,v)}^{(f_1,f_2)} \geq q_{p,(s,d)}^{f_1,u} + q_{p,(s,d)}^{f_2,v} - 1 \tag{19}$$

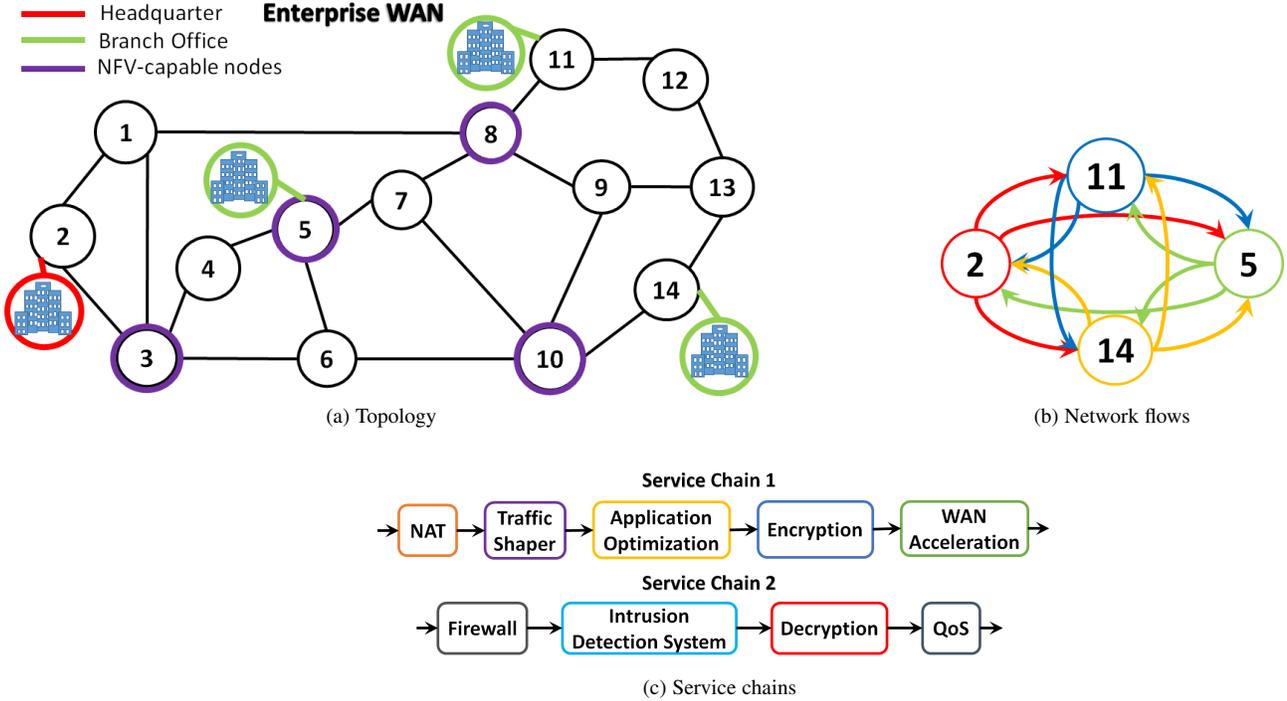

(a) Topology

(b) Network flows

**Service Chain 1**

**Service Chain 2**

(c) Service chains

Fig. 4: NSFNet topology and scenarios.

security, and then a QoS analysis is performed. Our illustrative examples only show results for SC1, as similar conclusions were also found for SC2.

For the DC-NFV case, we selected a set of four nodes (3, 5, 8, 10) as NFV-capable nodes (see Fig. 4(a)) on the basis of their nodal degree (nodes 3, 8, 10 have highest degree) and central location with respect to traffic flows (node 5)[7] Simulations are repeated and averaged for all possible locations of the DC and for all the service chaining strategies (see Section 4). For MB case, we also place MBs at the same four nodes (3, 5, 8, 10), but a single node can contain at a maximum 3 MBs for reliability and complexity purposes [28] [29]. Note that, since SC1 consists of 5 MBs, we require at least two nodes (i.e., one containing 3 MBs and the other 2 MBs) to be traversed to satisfy the service requirements. Simulations are then repeated for different combinations of placing MBs across the possible locations. We assume that MBs have sufficient processing capacity for the traffic in the network to avoid infeasible scenarios. Finally, for NFV ALL case, as all nodes can host VNFs, we run a single simulation for each combination of core count and traffic.

With the above simulation setting, we ran 7500 simulation experiments across all the service-chaining strategies, over an 8-core x86-64 bit Intel(R) Core(TM) i7-4770 CPU @ 3.40 GHz. Each simulation takes on average five minutes to be solved.

## 6.1 Results

Figures 5(a), 5(b), 5(c), 5(d), and 5(e) show the normalized **network resource consumption** $\varpi$ for increasing **core count** ($\theta$=2, 4, 8, 12, 24, 48, 96, and 192 CPU cores) in NFV-capable nodes, and for average traffic flows of 1 Gbps, 2.5 Gbps, 5 Gbps,

7.5 Gbps, and 10 Gbps, respectively. $\varpi$ for all the strategies is normalized to the $\varpi$ of "DC NFV ALL" strategy, which is used as a benchmark as it always returns the minimum possible value of $\varpi$[8].

As expected, given its scarce flexibility in service chaining, "MB" strategy has the highest $\varpi$ among all strategies and for any core count and traffic value. In MB case, network functions are statically placed in two nodes that must be reached by the traffic flows, and this causes longer paths and inefficient utilization of network resources. $\varpi$ decreases for "DC only" strategy. Also, in this case, network functions are statically located, but now traffic has to be routed only through a single node location (the DC), and not two fixed node locations (the MBs). Since "MB" does not depend on $\theta$, $\varpi$ remains constant for a given traffic value. The value for "MB" is cut-off at 1.6 as our discussion focuses on the other strategies.

Further decrease of $\varpi$ with respect to "DC only" is achieved using the "DC NFV x" strategies. While all of them implement NFV, the differences in $\varpi$ among these strategies result from the degree of flexibility in deploying SC instances. "DC only" can deploy multiple SC instances but all within the DC, while "DC NFV x" strategies can deploy SC instances across "x" NFV-capable nodes. Deploying SC instances across NFV-capable nodes enables traffic flows to be routed over shorter paths, thus significantly reducing $\varpi$. As expected, "DC NFV ALL" gives the minimum $\varpi$. Also, we find that, for the "DC NFV x" strategy, number of NFV-capable nodes, $\theta$, and traffic are the factors that affect $\varpi$. Let us analyze the effect of these factors in more detail.

---

7. NFV-capable node for "DC NFV 1" can be chosen in $\binom{4}{1}$ ways. Similarly, for "DC NFV 2", NFV-capable node selection can be done in $\binom{4}{2}$ ways.

8. Note that we do not consider "NFV ALL" strategy as a normalization standard as it becomes infeasible for certain small values of core count and large traffic. "NFV ALL" becomes infeasible because it does not feature any DC, while this does not occur with "DC NFV ALL" as the traffic can always be routed to the DC.

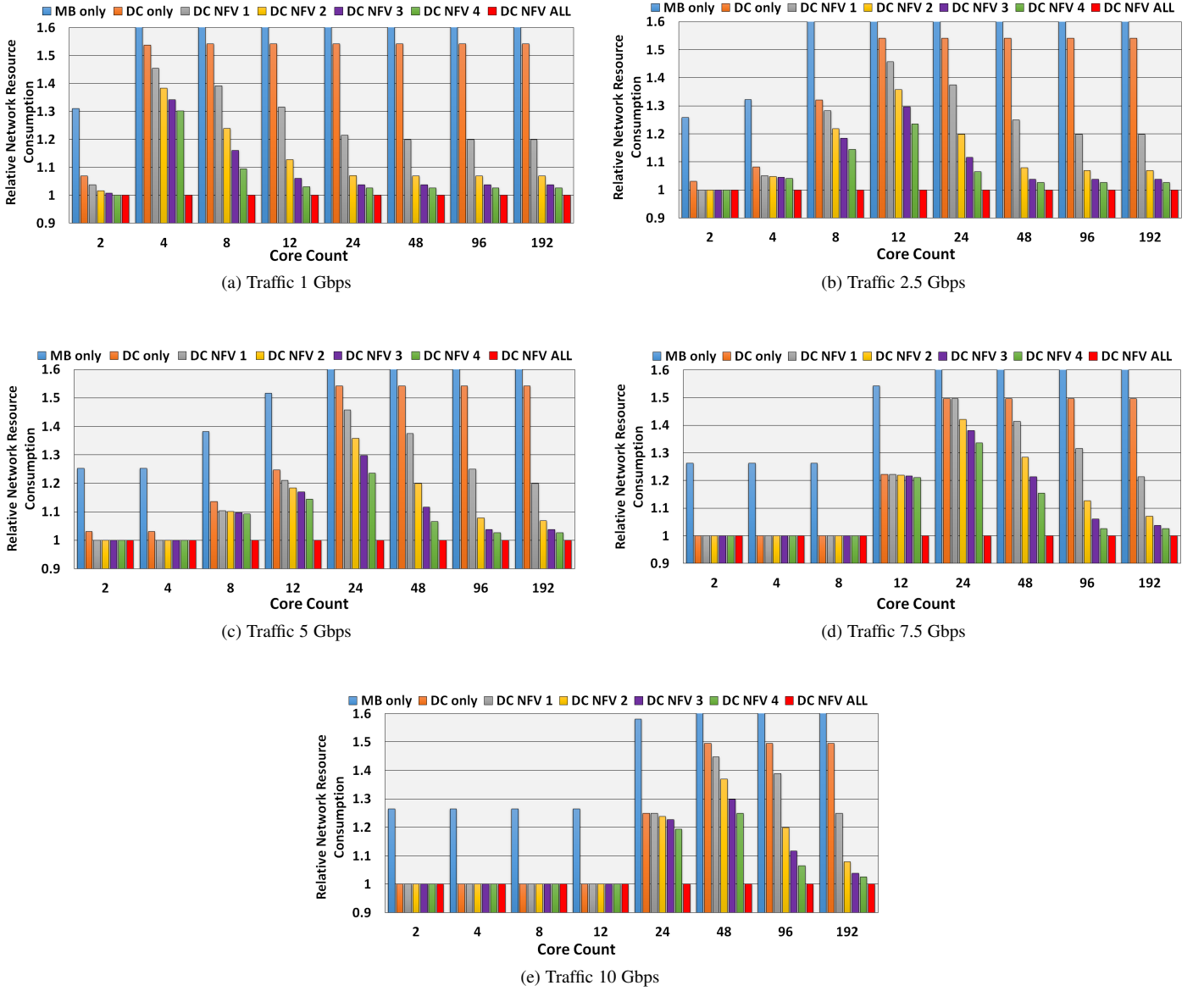

(a) Traffic 1 Gbps

(b) Traffic 2.5 Gbps

(c) Traffic 5 Gbps

(d) Traffic 7.5 Gbps

(e) Traffic 10 Gbps

Fig. 5: Relative network-resource consumption for different traffic values.

In Fig. 5(a), $\varpi$ does not vary much across the "DC NFV x" strategies for $\theta = 2$; but, at $\theta = 4$, a significant variation in $\varpi$ is observed. The reason for this variation is that "DC NFV ALL" at $\theta = 2$ is not able to effectively handle the traffic load inside the NFV-capable nodes as it does at $\theta = 4$, as $\theta = 2$ cannot pack enough VNFs as at $\theta = 4$. The variation in $\varpi$ at $\theta = 4$ between "DC NFV 1", "DC NFV 2", "DC NFV 3", and "DC NFV 4" occurs as more number of NFV-capable nodes are able to effectively handle traffic load since there are more CPU cores available, showing the effect the number of NFV-capable nodes have on $\varpi$. When we compare $\varpi$ values for $\theta = 4$ in Figs. 5(a) and 5(b), we find that $\theta = 4$, in the case of 5 Gbps average flow requests, is no longer able to effectively handle the traffic load in Fig. 5(b) because traffic load has increased. This same observation holds when we compare Figs. 5(a),5(b); 5(b),5(c); 5(c),5(d); and 5(d),5(e). In Figs. 5(d) and 5(e) at $\theta = 2, 4, 8$, we find that network-resource consumption of "DC only" strategy equals "DC NFV ALL". This is due to the fact that, for small core count, all traffic goes to the DC; hence, "DC NFV ALL" equals "DC only" at $\theta = 2, 4, 8$.

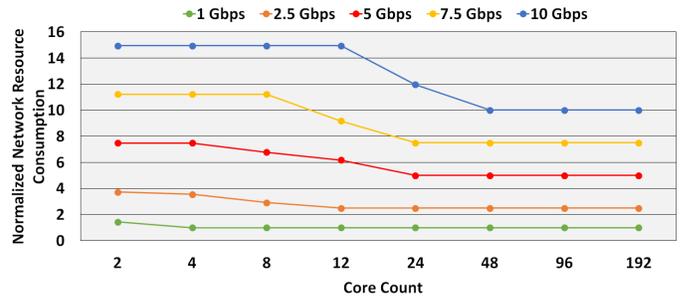

Fig. 6: DC NFV ALL.

Figure 6 shows $\varpi$ variation for "DC NFV ALL" across $\theta$ and for average traffic flows of 1 Gbps, 2.5 Gbps, 5 Gbps, 7.5 Gbps, and 10 Gbps. We plot "DC NFV ALL" as it is our best strategy, giving the minimum resource consumption across all scenarios. For ease of comparison, we normalize against $\varpi$ value for 1 Gbps traffic and 192 CPU cores. We find that $\theta$ value at which $\varpi$ becomes constant (meaning that most flows are served through shortest paths) is variable and depends on traffic load and core count. We call these combinations of core count and traffic load as "inflection points", corresponding to those values where "DC NFV ALL" has enough CPU cores to handle the traffic load with minimum resource consumption.

Across Figs. 5(a), 5(b), 5(c), 5(d), and 5(e), we find that, for values before the inflection points, $\varpi$ for "DC NFV x" increases as much as 10% w.r.t. "DC NFV ALL", and this happens as "DC NFV ALL" is able to handle service load more effectively than "DC NFV x". Based on this observation, we conclude that, after determining the inflection point (given the traffic scenario), an operator can select the number of nodes and upgrade them (i.e., install CPU cores) to get $\varpi$ within a desired percentage of the minimal. For example, "DC NFV 4" gives $\varpi$ with 10% of the minimum ("DC NFV ALL") beyond the "inflection points". Also, even at these inflection points, "DC NFV 4" utilizes 20% less $\varpi$ than "DC only" and close to 50% less at higher values of $\theta$ (beyond inflection point).

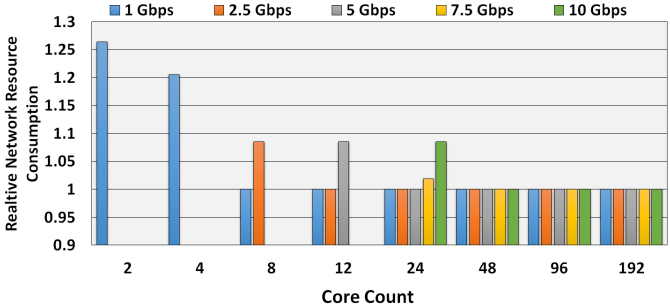

Fig. 7: NFV ALL vs. DC NFV ALL.

In Fig. 7, we plot network-resource consumption of "NFV ALL" relative to "DC NFV ALL". We see that "NFV ALL" is not feasible for all traffic values at lower $\theta$; but, as $\theta$ increases, service-chaining requirements can be satisfied and solution becomes feasible. Infeasible solutions occur when the $\theta$ required for processing that amount of traffic is not available among the NFV-capable nodes[9]. This result demonstrates that a centralized infrastructure like a DC is essential to the concept of a NeC, i.e., in some cases, an NeC cannot be realized as a fully-distributed infrastructure. At $\theta = 2$, "NFV ALL" manages service load for 1 Gbps traffic and becomes infeasible for higher traffic values. At $\theta = 8$, "NFV ALL" gives optimal $\varpi$ for 1 Gbps traffic, same as "DC NFV ALL" (inflection point at $\theta = 4$). Finally, at $\theta = 24$, "NFV ALL" can manage service load for all the traffic which was possible for "DC NFV ALL" at all $\theta$. Thus, we infer that a NeC must include centralized infrastructure (e.g., DC).

However, a centralized-only infrastructure will be prone to congestion at high traffic loads. In our mathematical formulation, congestion will occur when the **total incoming flow** ($\rho$) exceeds

9. It is an interesting problem on how to split a single VNF over two nodes in order to acquire the required CPU cores to process the given traffic. This problem is, however, out of scope for our present work and is an open problem for future research. In this work, we assume that a single VNF instance is completely contained within one node.

the **flow intake capacity** ($\Omega$) of a node, in which case the ILP will become infeasible for that node. Since we assume that all links in the network are bidirectional and have the same capacity, $\Omega$ for a node becomes product of the nodal degree and link capacity.

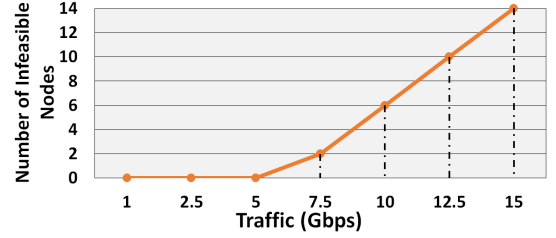

Fig. 8: Congestion trend in DC-only case.

In our considerations, "DC only" strategy represents centralized-only infrastructure. We run the "DC only" strategy for different traffic loads; and as earlier, different DC node placements ($\binom{14}{1}$ ways of choosing a DC node). Fig. 8 shows the number of nodes in the network that become infeasible (for DC placement) with increasing traffic. We find that, at 15 Gbps, all nodes in NSFNet become infeasible for DC placement. So, the "DC only" strategy becomes infeasible at $\geq$15 Gbps. We call this point the **congestion point** for the network.

| Traffic (Gbps) | $\rho$ (Gbps) | Infeasible DC nodes |
|---|---|---|
| 7.5 | 90 | 4,12 |
| 10 | 120 | 1,2,11,14 |
| 12.5 | 150 | 6,7,9,13 |
| 15 | 180 | 3,5,8,10 |

TABLE 2: Infeasible DC placements at different traffic values.

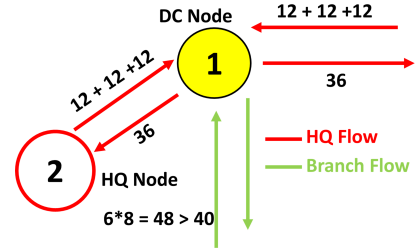

Fig. 9: Congestion in DC node 1 at 10 Gbps.

Table 2 shows which nodes become infeasible at what traffic values. As discussed earlier, if $\rho > \Omega$, then a DC node placement becomes infeasible. $\rho$ is the total flow in the network and is given as the product of average traffic and the total number of flows. This happens as all flows are required to traverse the DC node to satisfy their service requirements. If service needs of all traffic flows are not fulfilled, then the scenario becomes infeasible. The $\rho > \Omega$ relationship holds for all nodes except 1, 2, 5, 11 and 14. Since nodes 2, 5, 11, and 14 are source and sinks for all flows in the network, flows do not need to traverse them (6 of the 12 flows originate or terminate at nodes 2, 5, 11, or 14), and hence, they become infeasible at higher traffic than estimated by their $\Omega$. The anomaly of node 1 becoming infeasible at 10 Gbps occurs as result of non-uniform and unsplittable flows, one-hop distance from HQ node, traffic matrix, and network topology.

Fig. 9 shows the total incoming and outgoing flows into DC node 1. Though the average traffic in the network is 10 Gbps, there are six HQ flows of 12 Gbps each (all flows going to and from the HQ) and six branch flows of 8 Gbps each. Since node 2 is the HQ, it is source for three 12-Gbps flows and destination for three 12-Gbps flows. The HQ node is one hop away from the DC and cannot pack its flows with smaller flows because of the nature of the traffic matrix and the network topology. As a result, DC node 1 becomes infeasible since at least one traffic flow will not fulfill its service requirements.

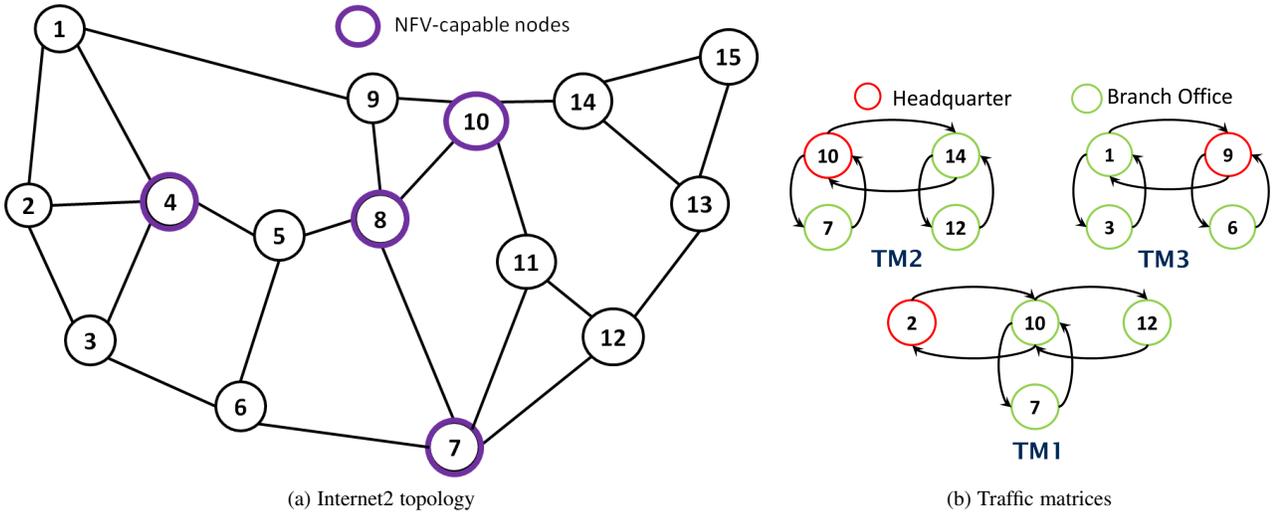

(a) Internet2 topology

(b) Traffic matrices

Fig. 10: Internet2 topology and traffic matrices.

The discussion on Figs. 7-9 makes the case for a Network-enabled Cloud ("DC NFV x"). In an NeC, we will not face resource scarcity like in a completely-distributed scenario ("NFV ALL") or congestion arising due to a single centralized infrastructure ("DC only"). This happens as we will have NFV-capable nodes along with centralized infrastructure like a DC. Also, Fig. 5 shows that having more NFV-capable nodes reduces $\varpi$ as it results in shorter flow paths. Additionally, an NeC scenario helps better balance the load on the links than in "DC only", where the links to/from the DC node get highly loaded. Below, we further analyze the congestion characteristics in the "DC only" strategy using different traffic matrices on Internet2 and GÉANT network topologies, and infer its effect on the network-resource consumption ($\varpi$) and load-balancing characteristics across strategies.

### 6.2 Congestion and Load Balancing

To study the robustness of our strategies and further verify our conclusions from previous results, we employ larger network topologies such as Internet2 and GÉANT, and study their characteristics.

#### 6.2.1 Internet2

Figure 10(a) shows the Internet2 [30] topology. All links in the network are bidirectional with capacity 40 Gbps each. We model an Enterprise WAN scenario where traffic matrices are as shown in Fig. 10(b). In each of TM1, TM2, and TM3, flows going to and from the HQ (Headquarter) carry 1.5 times the traffic between the branch offices. TM1 has two HQ flows and is a nation-wide Enterprise WAN, while TM2 and TM3 are eastern and western Enterprise WANs that have four HQ flows each. In this subsection, traffic values shown are average of the total flow in the network. All traffic flows across the traffic matrix (TM1, TM2, TM3) require SC1 (Fig. 4(c)) service. We consider only one of traffic in the network at any instance, i.e., all flows have the same service requirement.

NFV-capable nodes are selected based on nodal degree (4,7,8,10) and central location. Simulations across Figs. 11 and 12 are repeated and averaged for all possible locations of the DC.

We first analyze the congestion trend in Internet2 for different traffic matrices. Fig. 11 shows that the congestion points for TM3, TM1, and TM2 are 24 Gbps, 34 Gbps, and 37 Gbps, respectively, indicating that the traffic matrix affects the congestion trend in the network. From the discussion in the previous section, we know that congestion occurs when $\rho > \Omega$, while other factors such as nodes that are source-destination (s-d) pairs of network flows also contribute in shaping the congestion trend.

Table 3 shows that nodes 13 and 14 become infeasible as DC nodes for TM1 and TM3 at 14 Gbps even though $\rho(84) < \Omega(120)$. However, on analysis, nodes 13 and 14 can receive incoming traffic only from 14 and 13, which leads to $\Omega = 80$. So, $\rho(84) > \Omega(80)$ holds and the nodes become infeasible for DC placement. In TM2, however, 14 is feasible since it is one of the s-d pairs of TM2.

At 21 Gbps, node 12, though part of a s-d pair in TM1 and TM2, becomes infeasible. Similarly, node 3 in TM3 (part of a s-d

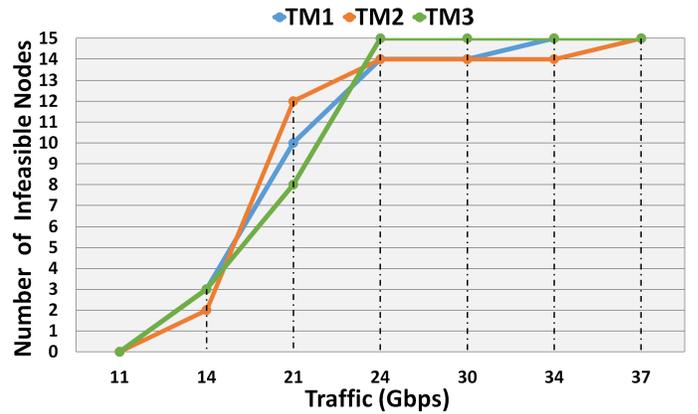

Fig. 11: Congestion trend in Internet2.

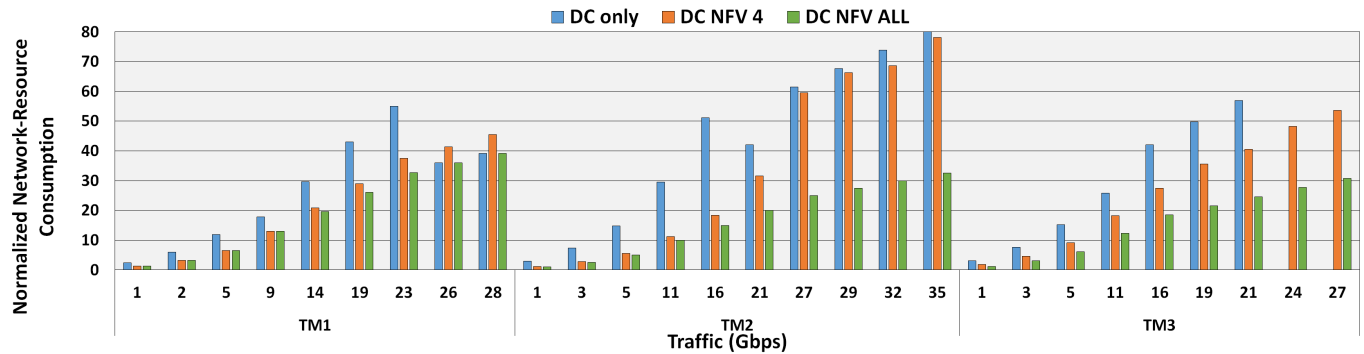

(a) Network-resource consumption in Internet2 across different strategies

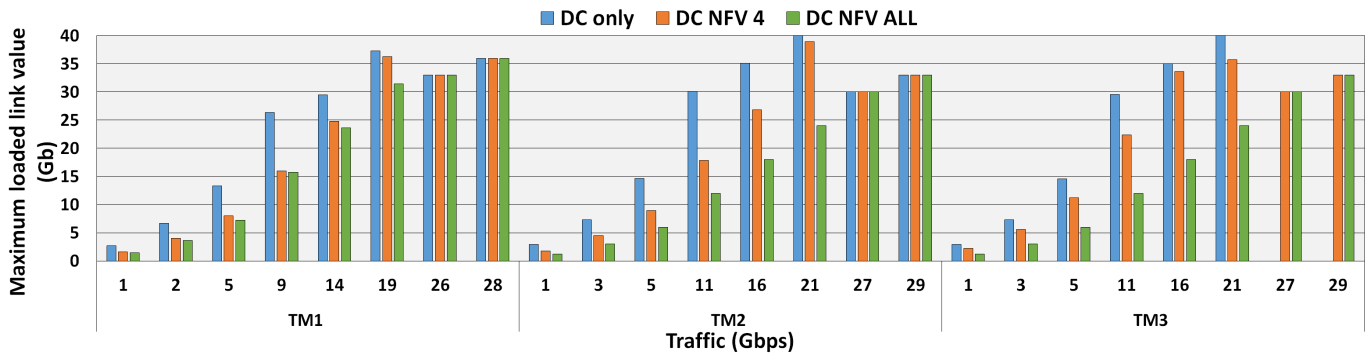

(b) Load balancing in Internet2

Fig. 12: Network-resource($\varpi$) consumption and load balancing in Internet2 topology.

TABLE 3: Congested nodes for DC placement in Internet2.

| Traffic (Gbps) | Infeasible DC nodes | | |
|---|---|---|---|
| | **TM1** | **TM2** | **TM3** |
| 14 | 13,14,15 | 13,15 | 13,14,15 |
| 21 | 1,3,5,6,9,11,12 | 1,2,3,4,5,6,9,11,12,14 | 2,3,5,11,12 |
| 24 | 2,4,7,8 | 7,8 | 1,4,6,7,8,9,10 |
| 34 | 10 | - | Infeasible |
| 37 | Infeasible | 10 | Infeasible |

TABLE 4: HQ and Branch (BR) flows for various traffic matrices. All values in Gbps.

| Traffic | $\rho$ (Total Flow) | TM1 | | TM2, TM3 | |
|---|---|---|---|---|---|
| | | HQ(2) | BR(4) | HQ(4) | BR(2) |
| 14 | 84 | 18 | 12 | 15.75 | 10.5 |
| 21 | 126 | 27 | 18 | 23.63 | 15.75 |
| 24 | 144 | 30.8 | 20.57 | 27 | 18 |
| 34 | 204 | 43.71 | 29.14 | 38.25 | 25.5 |

pair) and node 4 in TM2 ($\rho(126) < \Omega(160)$) also become infeasible though $\rho < \Omega$ has not been violated. This anomaly can be attributed to the combined effect of the network topology and traffic matrix, e.g., node 4 is feasible at 21 Gbps for TM3. This happens as (3,1) and (1,3) traffic flows can utilize the links associated with node 2. But, when looking at TM2, concentrated in the eastern US, we have to route the traffic to node 4, making links associated with node 2 obsolete and effectively reducing the nodal degree to 3 ($\Omega = 120$). So, we find that the network topology and traffic matrices are significant factors affecting the congestion trend.

TM3 reaches congestion point at 24 Gbps. Table 4 shows that, at 24 Gbps, HQ flows and branch (BR) flows are 27 and 18 Gbps, respectively. TM3 has four HQ flows and two BR flows. Since the flows are unsplittable, BR flows can be packed only with each other. If node 9 is the DC, it has two outgoing HQ flows, two incoming

HQ flows, and two incoming BR flows. It will not be possible to pack the two BR flows from (1,3) and (3,1) together. This would only be possible if the nodal degree of the node is four and shows how the flow values can affect the congestion trend.

The congestion point for TM1 is reached at 34 Gbps, while, for TM2, it is at 37 Gbps. The reason for this is the load of the flows in TM1 and TM2. TM1 has fewer HQ flows; and for it to have the same $\rho$ as TM2, its BR flows must have higher value. As shown in Table 4, this higher BR flow value leads to a infeasible HQ flow, and congestion is caused in TM1 showing as before that high flow values affect congestion in the network.

We find that a number of factors affect congestion in the case of a centralized infrastructure (i.e., DC) such as traffic matrix, network topology, unsplittable/splittable nature of traffic flows, and variation in flow value within a traffic matrix. Thus, $\rho > \Omega$ does not give a complete insight into congestion. This makes centralized infrastructure like a DC prone to congestion. Hence, though a DC has large amount of compute resources, NeC ("DC NFV x") deployment for NFV is a better alternative, as congestion in the network can be avoided.

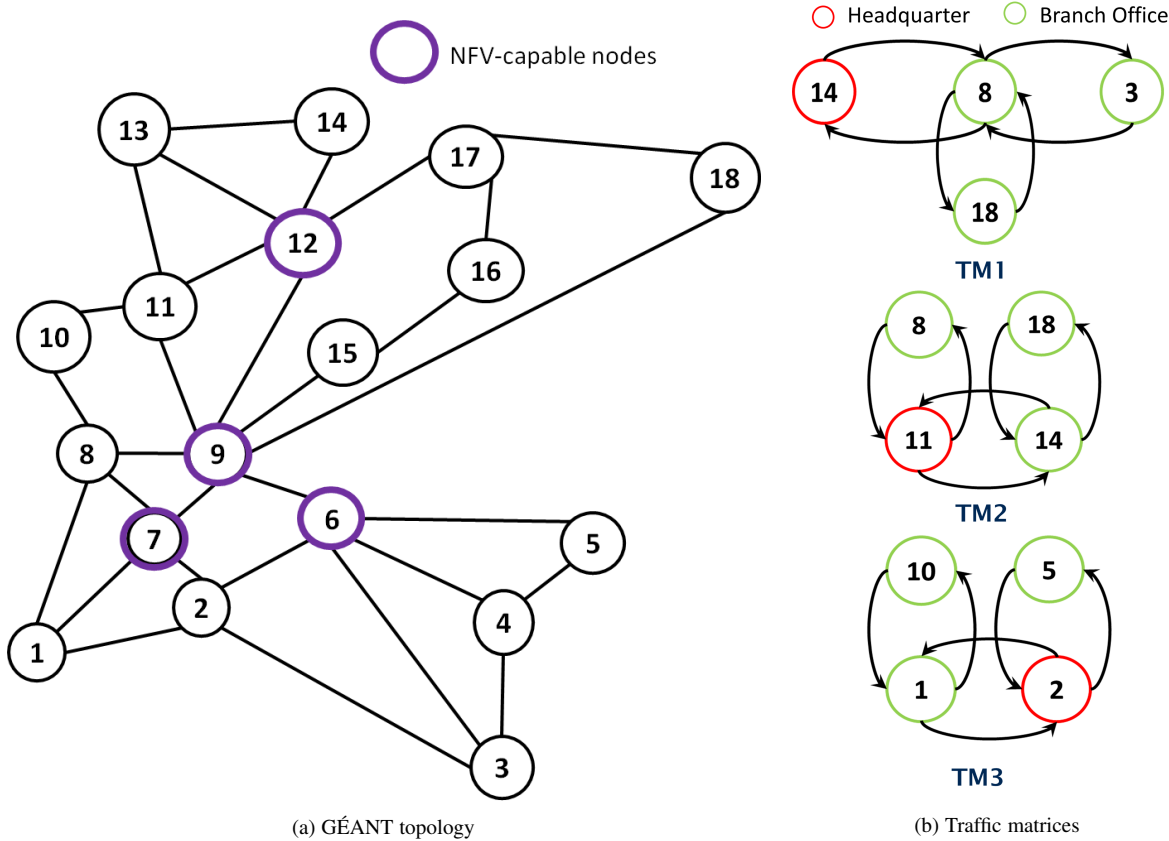

(a) GÉANT topology

(b) Traffic matrices

Fig. 13: GÉANT topology and traffic matrices.

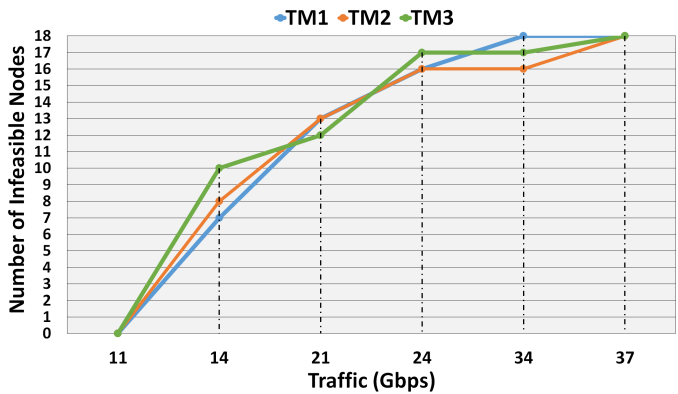

Fig. 14: Congestion trend in GÉANT.

The congestion trends indicate that, at higher traffic values, flows follow mutually-exclusive paths as they cannnot be aggregated with other flows. This explains the variation of the maximum loaded link value in Fig. 12(b) and exhibits the load-balancing property of our strategies. At higher loads, each flow occupies a single link; so, the maximum loaded link value remains the same across "DC only", "DC NFV 4", and "DC NFV ALL". "DC only" becomes infeasible in TM3 for 24 Gbps and higher as it gets congested at 24 Gbps as shown in Table 3.

Figure 12(a) shows the network-resource consumption ($\varpi$) for "DC only", "DC NFV 4", and "DC NFV ALL" for the various traffic matrices. Here, we keep $\theta = 192$ for "DC NFV 4" and "DC NFV ALL" to have enough compute resources even at larger loads. All values in Fig. 12(a) have been normalized against the value for "DC NFV ALL" at 1 Gbps for TM2.

In TM1, we find that "DC NFV 4" performs as well as "DC NFV ALL" until 10 Gbps. This happens as TM1 is a nation-wide Enterprise WAN, and all the flows have to traverse the central part of the network where NFV-capable nodes are situated. At higher traffic values of 26 and 28 Gbps, only node 10 is feasible as the "DC" node for "DC only" while, for "DC NFV 4", all DC nodes are still feasible since all flows do not go to the DC. Thus, the average network-resource consumption ($\varpi$) across the various DC placements for "DC NFV 4" becomes higher. Again, node 10's central nature and TM1's nation-wide nature causes the "DC NFV ALL" and "DC only" to have the same $\varpi$.

Across TM1, TM2, and TM3, we find that TM2 and TM3 have higher $\varpi$ for "DC only" and "DC NFV 4" since TM2 and TM3 are not nation-wide. But $\varpi$ is smaller for "DC NFV ALL" for TM2 and TM3 than TM1 since the flow paths are shorter because of the coverage area of TM2 and TM3 and all nodes are NFV-capable.

The trend displayed across Fig. 12 confirms the reduction of network-resource consumption and congestion by NeC ("DC NFV 4", "DC NFV ALL") with added benefit of load balancing.

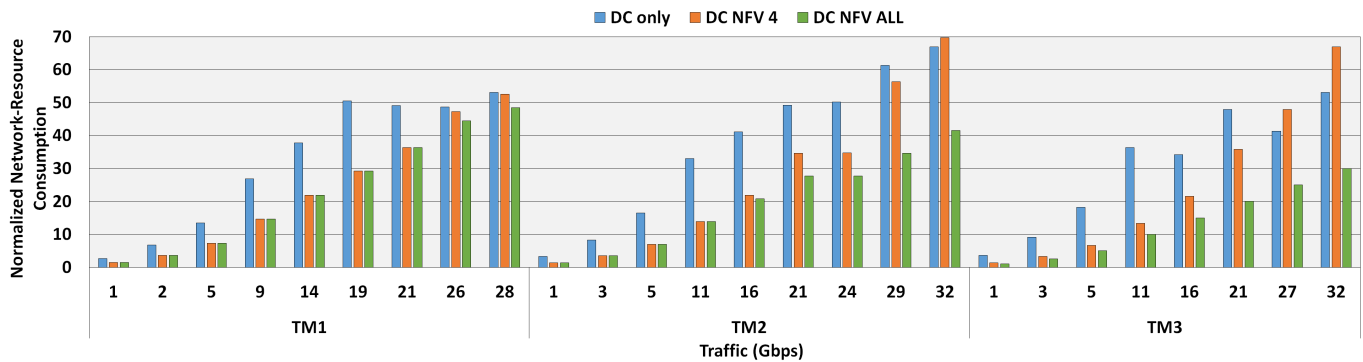

Fig. 15: Normalized network-resource consumption in GÉANT across different strategies.

### 6.2.2  GÉANT

The GÉANT [31] network topology is shown in Fig. 13(a). All links in the network are bidirectional with capacity 40 Gbps each. An Enterprise WAN scenario is modeled with traffic matrices TM1, TM2, and TM3. TM1 is a Europe-wide Enterprise WAN while TM2 and TM3 are restricted to northern and southern Europe, respectively. Here too, the HQ (Headquarter) carries 1.5 times the traffic between the branch offices. TM1 has two HQ flows while TM2 and TM3 have four HQ flows each. Only one type of traffic for one traffic matrix (TM1, TM2, or TM3) is considered at any instance, and it is the SC1 (Fig. 4(c)) service requirement traffic.

NFV-capable nodes (6,7,9,12) are chosen for their nodal degree and central nature. All results showing "DC NFV x" strategies had $\theta = 192$. This $\theta$ value avoids infeasibility at high traffic loads.

The congestion trend is shown in Fig. 14; and, as before, we find that the congestion point varies between traffic matrices. This again makes the case for a NeC to avoid congestion in NFV deployment.

Similarly, we find that $\varpi$ (Fig. 15) also exhibits trends explained in Fig. 12(a) showing that the traffic matrix affects $\varpi$ in different ways across strategies. All values in Fig. 15 have been normalized with $\varpi$ for "DC NFV ALL" at 1 Gbps for TM3.

### 6.3  Memory-Based VNF

| Application | Throughput | | |
|---|---|---|---|
| | 5 Gbps | 10 Gbps | 20 Gbps |
| NAT | 1 GB | 2 GB | 4 GB |
| IPsec VPN | 1 GB | 2 GB | 4 GB |
| Application Optimization | 2 GB | 4 GB | 8 GB |

TABLE 5: Memory requirements for various applications [32] [33].

In Section 3, we elaborated on the CPU-core-to-throughput relationships of VNFs. Here, we analyze the scenario where VNFs are more dependent on memory than CPU.

Table 5 shows memory requirements of middle-boxes that have been configured for a particular application. For lack of available information on actual VNF memory requirements (as far as we are aware of), the table is used to estimate requirements for VNFs. However, memory-consumption characteristics of memory-intensive VNFs are not apparent, i.e., memory requirements may

scale with increase in throughput (scaling) or remain static (non-scaling) as follows:

- **Non-Scaling VNF (NS)**: Memory requirements are for installation; operation does not require memory scaling.
- **Scaling VNF (S)**: These VNFs require more memory as throughput demand increases.

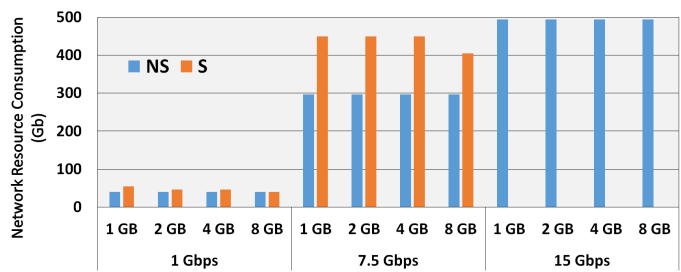

Fig. 16: Resource usage in DC NFV 4 for NS and S type VNFs.

We run our simulation on the "DC NFV 4" strategy for SC1 on the scenario mentioned in Section 6 and Fig. 4 using NSFNet. To deploy memory-intensive VNFs, we relax Eq. (4) and alternatively, enforce Eqs. (5) and (6) for "S" and "NS" VNFs, respectively.

Fig. 16 reaffirms that network-resource consumption ($\varpi$) for "NS" is not dependent on available memory while "S" type VNFs require memory scaling, and, hence, have lower $\varpi$ values as available memory increases. At 15 Gbps, we find that "S" becomes infeasible as the available memory does not satisfy service needs and congestion point (as explained in Fig. 8) for NSFNet is reached.

## 7  CONCLUSION

We introduced the problem of service chaining and how Virtual Network Functions (VNFs) can be used for more flexible/agile service chaining in a Network-enabled Cloud (NeC). We defined and analyzed different service-chaining strategies for different traffic demand values and different computing core capacity using an Integer Linear Program (ILP). Our objective was to demonstrate the network-resource reduction that can be achieved by employing the concept of a NeC with a DC node and a certain number of NFV-capable nodes. We studied multiple strategies under the NeC concept and compared them with the traditional middle-box and DC strategies. We found that, by determining the right combination of core count and offered traffic and selectively upgrading a set of nodes to support NFV technology, we can achieve close-to-optimal network-resource consumption while significantly reducing network-resource consumption in comparison to a centralized infrastructure such as a DC-only solution.

We also studied the congestion trend in various network topologies for multiple traffic matrices when deploying centralized infrastructure like a DC. We demonstrated that the NeC approach for NFV deployment results in reduction of congestion, reduced network-resource consumption, and reduced link loads than a centralized infrastructure.

## ACKNOWLEDGMENT

This work was supported by NSF Grant No. CNS-1217978.

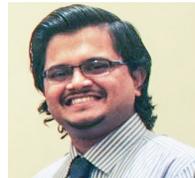

**Abhishek Gupta** (abgupta@ucdavis.edu) received his Bachelor of Technology (B.Tech) degree in computer science and engineering from Vellore Institute of Technology, India in 2012. He is currently a Ph.D. candidate at the University of California, Davis. His research interests include green and cost-effective cloud infrastructures and Network Function Virtualization (NFV). He is a student member of IEEE and is a co-recipient of best paper awards at IEEE ANTS 2014 and 2015 conferences.

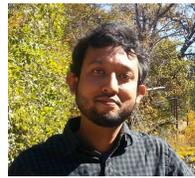

**M. Farhan Habib** is currently a Software Engineer at Intel Corp. in Folsom, California. He received his PhD from University of California, Davis in 2014 with a GGCS Best Graduate Researcher Honorable Mention. His research interests include survivability, programmability, and energy efficiency of communication networks, and compiler optimization.

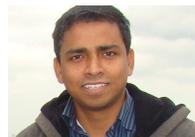

**Uttam Mandal** is a Senior Software Engineer at Cisco Systems Inc. in San Jose, California. He received his PhD and M.S. degree from University of California, Davis in computer science, and bachelor of engineering (B.E.) from Javadpur University, Kolkata, India. He authored several peer-reviewed scholarly papers in international journal and conference proceedings. His research focused on resource-efficient cloud and content delivery networking systems design. His research also includes design of energy-efficient and green-energy-aware networking systems. Currently he is working in designing and developing Network Function Virtualization (NFV) and Orchestration for service provider networks. His previous experience includes content (video) delivery systems for industry-scale content owners and broadcasters.

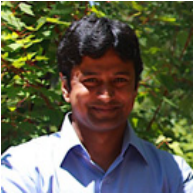

**Pulak Chowdhury** (pchowdhury@ucdavis.edu) is an author of 30+ peer-reviewed scholarly papers in international journal and conference proceedings, mostly in resource-efficient, next-generation networking domain. He was an associate editor of IEEE Journal of Selected Areas in Communications (JSAC), Special Issue on Emerging Technologies in Communications. He has also served as a Technical Program Committee (TPC) member of international conferences and workshops. Dr. Chowdhury received the Ph.D. degree from University of California, Davis, the M.A.Sc. degree from McMaster University, Canada, and the B.Sc. Engineering degree from Bangladesh University of Engineering and Technology, Dhaka, Bangladesh, in 2011, 2005, and 2002, respectively. His research interests cover a variety of topics in optical networks, hybrid wireless-optical networks, and energy efficiency in next-generation networks. He is Founder and Chief Architect at Ennetix inc.

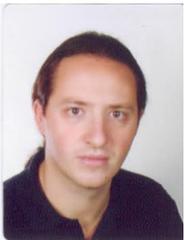

**Massimo Tornatore** is currently an Associate Professor in the Department of Electronics, Information and Bioengineering at Politecnico di Milano, Italy, where he received a PhD degree in Information Engineering in 2006. He also holds an appointment as adjunct associate professor in the Department of Computer Science at the University of California, Davis.

He is author of more than 200 peer-reviewed conference and journal papers and his research interests include performance evaluation, optimization and design of communication networks (with an emphasis on the application of optical networking technologies), cloud computing, and energy-efficient networking. He is member of the editorial board of Springer journal "Photonic Network Communications" and of the Technical Program Committee of various networking conferences such as INFOCOM, OFC, ICC, Globecom, etc. He is a senior member of the IEEE and he was a co-recipient of six best-paper awards from IEEE conferences.

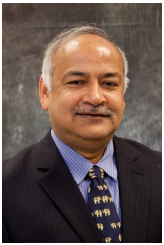

**Biswanath (Bis) Mukherjee** is a Distinguished Professor at University of California, Davis, where he has been a faculty member since 1987 and was Chairman of Computer Science during 1997-2000. He received the BTech (Hons) degree from Indian Institute of Technology, Kharagpur (1980) and PhD from University of Washington, Seattle (1987). He was General Co-Chair of the IEEE/OSA Optical Fiber Communications (OFC) Conference 2011, Technical Program Co-Chair of OFC2009, and Technical Program, Chair of the IEEE INFOCOM96 conference. He is Editor of Springers Optical Networks Book Series. He has served on eight journal editorial boards, most notably IEEE/ACM Transactions on Networking and IEEE Network. In addition, he has guest-edited Special Issues of Proceedings of the IEEE, IEEE/OSA Journal of Lightwave Technology, IEEE Journal on Selected Areas in Communications, and IEEE Communications. To date, he has supervised 66 PhDs to completion and currently mentors 15 advisees, mainly PhD students. He is winner of the 2004 Distinguished Graduate Mentoring Award and the 2009 College of Engineering Outstanding Senior Faculty Award at UC Davis. He is co-winner of 10 Best Paper Awards from various conferences, including Optical Networking Symposium Best Paper Awards at IEEE Globecom 2007 and 2008. He is author of the graduate-level textbook Optical WDM Networks (Springer, January 2006). He served a 5-year term on Board of Directors of IPLocks, a Silicon Valley startup company (acquired by Fortinet). He has served on Technical Advisory Board of several startup companies, including Teknovus (acquired by Broadcom). He is Founder and President of Ennetix, Inc., a startup company incubated at UC Davis and developing cloud-based network performance analytics and management software. He is an IEEE Fellow.